\documentclass[sigconf,table,xcdraw]{acmart}

\AtBeginDocument{%
  }

\usepackage{multirow}
\usepackage{booktabs}
\usepackage{colortbl}
\usepackage{hhline}
\usepackage{algorithm}
\usepackage{algpseudocode}
\usepackage{graphicx}
\usepackage{subcaption}
\usepackage{array}
\usepackage{tikz}
\usepackage{pifont}
\usepackage{balance}

\newcommand{\emojiCheck}{\includegraphics[height=0.6em]{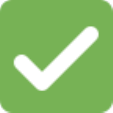}}
\newcommand{\emojiBox}{\includegraphics[height=0.6em]{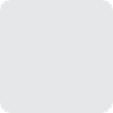}}

\hypersetup{
    colorlinks=true,     
    linkcolor=blue,       
    citecolor=blue,       
    filecolor=blue,       
    urlcolor=blue         
}

\copyrightyear{2025} 
\acmYear{2025} 
\setcopyright{acmlicensed}\acmConference[SIGIR '25]{Proceedings of the 48th International ACM SIGIR Conference on Research and Development in Information Retrieval}{July 13--18, 2025}{Padua, Italy}
\acmBooktitle{Proceedings of the 48th International ACM SIGIR Conference on Research and Development in Information Retrieval (SIGIR '25), July 13--18, 2025, Padua, Italy}
\acmDOI{10.1145/3726302.3730222}
\acmISBN{979-8-4007-1592-1/2025/07}

\begin{document}

\title[LLM-based Query Expansion Fails for Unfamiliar and Ambiguous Queries]{LLM-based Query Expansion Fails \\ for Unfamiliar and Ambiguous Queries}

\author{Kenya Abe}
\authornote{Currently with NEC Corporation.}
\email{s1911448@klis.tsukuba.ac.jp}
\orcid{0009-0007-5413-3158}
\affiliation{%
  \institution{University of Tsukuba}
  \city{Tsukuba}
  \country{Japan}
}

\author{Kunihiro Takeoka}
\email{k_takeoka@nec.com}
\orcid{0009-0000-9653-8523}
\affiliation{%
  \institution{NEC Corporation}
  \city{Tokyo}
  \country{Japan}
}

\author{Makoto P. Kato}
\email{mpkato@acm.org}
\orcid{0000-0002-9351-0901}
\affiliation{%
  \institution{University of Tsukuba}
  \city{Tsukuba}
  \country{Japan}
}
\additionalaffiliation{%
  \institution{National Institute of Informatics}
  \city{Tokyo}
  \country{Japan}
}

\author{Masafumi Oyamada}
\email{oyamada@nec.com}
\orcid{0000-0002-4045-7350}
\affiliation{%
  \institution{NEC Corporation}
  \city{Tokyo}
  \country{Japan}
}

\begin{abstract}
Query expansion (QE) enhances retrieval by incorporating relevant terms, with large language models (LLMs) offering an effective alternative to traditional rule-based and statistical methods. However, LLM-based QE suffers from a fundamental limitation: it often fails to generate relevant knowledge, degrading search performance. Prior studies have focused on hallucination, yet its underlying cause—LLM knowledge deficiencies—remains underexplored. This paper systematically examines two failure cases in LLM-based QE: (1) when the LLM lacks query knowledge, leading to incorrect expansions, and (2) when the query is ambiguous, causing biased refinements that narrow search coverage. We conduct controlled experiments across multiple datasets, evaluating the effects of knowledge and query ambiguity on retrieval performance using sparse and dense retrieval models.
Our results reveal that LLM-based QE can significantly degrade the retrieval effectiveness when knowledge in the LLM is insufficient or query ambiguity is high. We introduce a framework for evaluating QE under these conditions, providing insights into the limitations of LLM-based retrieval augmentation.
\end{abstract}

\ccsdesc[500]{Information systems~Query reformulation}

\keywords{Query Expansion, Large Language Models, Document Retrieval}

\maketitle

\section{Introduction}
Query expansion (QE)~\cite{robertson1976relevance,carpineto2012survey} is an established technique in information retrieval that improves search accuracy by adding terms relevant to the user's intent. 
This method addresses the inherent challenge that user queries often lack sufficient specificity to retrieve relevant results. 
Traditional QE approaches, including rule-based~\cite{pal2014improving} and statistical~\cite{cao2008selecting} techniques, refine queries using predefined term relationships or corpus-based co-occurrence statistics. 

Recent advances in large language models (LLMs), such as GPT and LLaMA series~\cite{openai2022chatgpt,meta2024llama3}, introduce an effective approach to QE. By leveraging their pre-trained knowledge, LLM-based QE methods~\cite{wang2023query2doc,mackie2023generative,young2024gaqr,jagerman2023query} generate expansions better aligned with user intent. However, a limitation remains: LLMs may fail to generate relevant knowledge for QE, degrading retrieval performance.

Previous studies have examined hallucination as a key challenge in LLM-based QE~\cite{lei2024corpus,baek2024crafting}. Hallucination refers to the generation of factually incorrect or inconsistent information to a user query~\cite{huang2025hallucination}, which in QE can introduce misleading or irrelevant terms, reducing retrieval effectiveness. However, hallucination in QE stems from more fundamental issues: LLMs may either lack the necessary knowledge to generate meaningful expansions or fail to utilize their knowledge effectively due to query ambiguity. Despite growing interest in LLM-based QE, existing research has not examined these failures systematically and thoroughly.

This study systematically investigates how LLMs fail to expand queries and how these failures impact retrieval effectiveness. We identify two primary failure scenarios: (1) When the LLM lacks knowledge of the query, it may introduce non-existent entities or unrelated terms, decreasing retrieval accuracy. (2) When the query is ambiguous, the LLM may generate biased expansions, which narrow the search coverage and exclude relevant documents. Understanding these failure cases is essential for determining when LLM-based QE enhances or degrades retrieval effectiveness.

To address these challenges, we investigate the following research questions:
\textbf{RQ1}: How does QE effectiveness differ between queries for which LLM has sufficient knowledge and those for which it does not? \textbf{RQ2}: How does query ambiguity affect the effectiveness of QE?
We conduct experiments on multiple datasets, systematically controlling knowledge availability and query ambiguity. 

Our results revealed that (1) the effectiveness of LLM-based QE degrades when the LLM is unfamiliar with the query, especially in general domains rather than specialized domains. (2) 
the QE effectiveness in terms of recall 
is lower for highly ambiguous queries in most datasets.
Furthermore, we observed that LLM-based QE was effective in retrieving relevant documents for popular query interpretations, but not for less popular ones.

\section{Related Work}
\label{sec:expansion_format}
LLM-based QE methods can be categorized into three categories according to Zhu et al.~\cite{zhu2023large}: (1) {\bf answer}, which generates pseudo-answers to a given query, as expansion terms~\cite{wang2023query2doc,jagerman2023query}; (2) {\bf keyword}, which produces relevant terms to a given query~\cite{jagerman2023query,mackie2023generative}; and (3) {\bf question}, which rewrites a given query in a question format~\cite{young2024gaqr,ma2023query}. 
In our experiments, we examined a representative method from each category, namely, 
Q2D~\cite{wang2023query2doc},
Q2E~\cite{jagerman2023query}, and GaQR~\cite{young2024gaqr}.

Previous research has explored factors that affect the effectiveness of LLM-based QE. Weller et al.~\cite{weller-etal-2024-generative} showed that QE can harm high-performing retrieval models, while Ayoub et al.~\cite{ayoub2024case} found answer-type expansions more effective in general-domain datasets. However, the relationship between LLM-specific challenges and QE remains underexplored.

Most studies on LLM knowledge gaps and hallucinations rely on limited examples or small-scale experiments~\cite{lei2024corpus,baek2024crafting}. To fill this gap, we conduct large-scale experiments using diverse datasets, retrieval models, and expansion methods, focusing on how the limitations of LLM knowledge affect QE. Our study offers a more comprehensive perspective than prior work limited to a single dataset or model.

While query ambiguity is widely recognized as a critical problem~\cite{zhang2024clamber,dai2024bias}, its specific impact on LLM-based QE is less studied~\cite{anand2023context,yu2023search}. Anand et al.~\cite{anand2023context} highlighted the risks of ambiguity in QE, and Yu et al.~\cite{yu2023search} explored generating diverse query facets. Our work provides an in-depth evaluation of how query ambiguity influences LLM-based QE performance.

\section{Experimental Settings}

\subsection{Datasets}\label{sec:dataset}

\textbf{To answer RQ1}, we divided the query set into two groups based on the LLM's knowledge of the query. We investigated how QE performance differs between these groups (see Section~\ref{sec:llm_evaluation} for details on LLM's knowledge evaluation). 
We selected the following four datasets.
\textbf{Natural Questions (NQ) and TriviaQA (TQ)}~\cite{kwiatkowski2019natural} contain questions that can often be answered with just a few words. For NQ and TQ, we used the data provided by KILT~\cite{petroni2020kilt} and evaluated the retrieval effectiveness under the same settings as Salemi et al.~\cite{salemi2024evaluating}.
\textbf{MS MARCO}~\cite{nguyen2016ms} contains web search queries, including non-factoid questions that cannot be answered with just a few words. \textbf{BioASQ}~\cite{tsatsaronis2015overview} is a specialized biomedical domain, unlike the general domain that dominates most LLMs' training data. 

\textbf{To answer RQ2}, we used the following three datasets, splitting the query set into two groups based on query ambiguity and comparing the results of these groups.
\textbf{AmbigDocs}~\cite{lee2024ambigdocs} and \textbf{AmbigQA}~\cite{min2020ambigqa} are QA datasets built on Wikipedia. Although both address question ambiguity, they differ in scope: AmbigQA covers a broad range of ambiguity types, whereas AmbigDocs focuses solely on the ambiguity of entities referred to in queries. 
Since AmbigQA does not provide relevant passages, we used the weak relevance labels provided by the AmbigQA authors, following previous research~\cite{prabhu2024dexter}. 
As the granularity of predefined interpretations of each query differs across the datasets, 
we set a different threshold for each dataset to determine low- and high-ambiguity queries.
In AmbigDocs, we classified queries with two or three interpretations as low-ambiguity queries and those with four or more interpretations as high-ambiguity queries. In AmbigQA, we defined queries with a single interpretation as low ambiguity and those with more than one interpretation as high ambiguity.
\textbf{WebTrack09-12 (TREC Web)}~\cite{clarke2009overview} comprises web search topics and web corpus~\cite{clueweb09}. We treated queries with four or more subtopics as high ambiguity, and those with three or fewer as low ambiguity. Due to concerns about the reusability of these test collections, we used {\it condensed-list}~\cite{sakai2013unreusability} that treats unjudged documents as not retrieved. Although condensed-list has the issue of overestimating systems that did not participate in pooling~\cite{sakai2007alternatives,sakai2013unreusability}, we believe this overestimation is not a critical problem in our experiments since our objective is to evaluate the results from the same retrieval model comparatively.

\subsection{Retrieval Components}

\subsubsection{Retrieval Models.} In this study, we used three first-stage retrieval models: BM25, Contriever~\cite{izacard2021unsupervised}, and E5-base-v2~\cite{wang2022text}. We set the BM25 hyperparameters to $k1=0.9$ and $b=0.4$, and input lengths of Contriever and E5-base-v2 to 256 for both queries and documents.

\subsubsection{LLMs for Query Expansion}
We used two different LLMs for QE, namely, 
gpt-3.5-turbo-0125 (GPT)~\cite{openai2022chatgpt} and Meta-Llama-3-8B-Instruct (Llama8b)~\cite{meta2024llama3}.
In our analysis, we used zero-shot prompting for all expansion models. 

\subsubsection{Query Expansion Methods}

Based on the categorization~\cite{zhu2023large} in Section~\ref{sec:expansion_format}, we selected three QE methods.
\textbf{Query2Expansion (Q2E)}~\cite{jagerman2023query} (\textbf{keyword} category) generates query-related keywords using an LLM.
\textbf{Query2Doc (Q2D)}~\cite{wang2023query2doc} (\textbf{answer} category) generates a pseudo-answer document for the query via an LLM.
\textbf{GaQR}~\cite{young2024gaqr} (\textbf{question} category) rewrites the query to improve search results.
GaQR is a QE model that requires training of an LLM. 
It trains a smaller LLM (student) using rewritten queries generated from a larger LLM (teacher).
From a fairness perspective, 
we opted for a different implementation than that of the original work.
We implemented GaQR using gpt-3.5-turbo-0125 as the teacher (T) and Meta-Llama-3-8B-Instruct as the student (S), both of which are employed in the other QE methods in our experiments.
We tuned the hyperparameters as a part of our effort to reproduce GaQR\footnote{In GaQR paper, Recall@100 and Recall@1000 were 65.6 and 87.1 on MS MARCO-dev, respectively, whereas our reproduction achieved better results: 69.2 and 88.5.}. 

\begin{table*}[t]
  \caption{
  Improvements from QE under two knowledge conditions: the LLM lacks sufficient knowledge (\emojiBox) vs. has sufficient knowledge(\emojiCheck). Bold values denote statistically significant differences between the two knowledge conditions. 
  }
  \centering
  \scalebox{0.77}{%
    \begin{tabular}{llcccccccccccccccc}
\toprule
                            &                             & \multicolumn{4}{c}{NQ}                                                                                                                                                            & \multicolumn{4}{c}{TQ}                                                                                                                                      & \multicolumn{4}{c}{MS MARCO}                                                                                                                                   & \multicolumn{4}{c}{BioASQ}                                                                                                                                    \\ \cmidrule(lr){3-6} \cmidrule(lr){7-10} \cmidrule(lr){11-14} \cmidrule(lr){15-18}  
\multirow{2}{*}{Retrieval}  & \multirow{2}{*}{Expansion}  & \multicolumn{2}{c}{NDCG@10}                                                             & \multicolumn{2}{c}{Recall@100}                                                          & \multicolumn{2}{c}{NDCG@10}                                        & \multicolumn{2}{c}{Recall@100}                                                          & \multicolumn{2}{c}{NDCG@10}                                         & \multicolumn{2}{c}{Recall@100}                                                           & \multicolumn{2}{c}{NDCG@10}                                                              & \multicolumn{2}{c}{Recall@100}                                     \\ \cmidrule{3-18} 
                            &                             & \raisebox{0.20ex}{\huge \emojiBox}                              & \raisebox{0.20ex}{\huge \emojiCheck}                                                        & \raisebox{0.20ex}{\huge \emojiBox}                              & \raisebox{0.20ex}{\huge \emojiCheck}                                                        & \raisebox{0.20ex}{\huge \emojiBox}                             & \raisebox{0.20ex}{\huge \emojiCheck}                                                      & \raisebox{0.20ex}{\huge \emojiBox}                              & \raisebox{0.20ex}{\huge \emojiCheck}                                                       & \raisebox{0.20ex}{\huge \emojiBox}                              & \raisebox{0.20ex}{\huge \emojiCheck}                                                        & \raisebox{0.20ex}{\huge \emojiBox}                              & \raisebox{0.20ex}{\huge \emojiCheck}                                                        & \raisebox{0.20ex}{\huge \emojiBox}                              & \raisebox{0.20ex}{\huge \emojiCheck}                                                       & \raisebox{0.20ex}{\huge \emojiBox}                              & \raisebox{0.20ex}{\huge \emojiCheck}                                  \\[-1ex] \midrule \midrule
\multirow{4}{*}{BM25}       & \multicolumn{1}{l|}{No Exp} & \cellcolor[HTML]{C0C0C0}21.69 & \multicolumn{1}{c|}{\cellcolor[HTML]{C0C0C0}19.94}      & \cellcolor[HTML]{C0C0C0}29.63 & \multicolumn{1}{c|}{\cellcolor[HTML]{C0C0C0}24.40}      & \cellcolor[HTML]{C0C0C0}22.07 & \multicolumn{1}{c|}{\cellcolor[HTML]{C0C0C0}22.45}      & \cellcolor[HTML]{C0C0C0}22.25 & \multicolumn{1}{c|}{\cellcolor[HTML]{C0C0C0}21.63}      & \cellcolor[HTML]{C0C0C0}27.24  & \multicolumn{1}{c|}{\cellcolor[HTML]{C0C0C0}20.05}      & \cellcolor[HTML]{C0C0C0}68.01  & \multicolumn{1}{c|}{\cellcolor[HTML]{C0C0C0}64.38}      & \cellcolor[HTML]{C0C0C0}57.36  & \multicolumn{1}{c|}{\cellcolor[HTML]{C0C0C0}39.66}      & \cellcolor[HTML]{C0C0C0}82.54 & \cellcolor[HTML]{C0C0C0}64.53      \\
                            & \multicolumn{1}{l|}{Q2E}    & +5.98 & \multicolumn{1}{c|}{\bf{+7.74}} & +6.31 & \multicolumn{1}{c|}{+6.54}      & +1.68 & \multicolumn{1}{c|}{\bf{+2.32}} & +1.81 & \multicolumn{1}{c|}{\bf{+2.52}} & -1.60  & \multicolumn{1}{c|}{\bf{+1.59}} & -0.75  & \multicolumn{1}{c|}{\bf{+4.47}} & +1.51  & \multicolumn{1}{c|}{+0.33}      & -0.16 & +3.19      \\
                            & \multicolumn{1}{l|}{Q2D}    & +7.42 & \multicolumn{1}{c|}{\bf{+9.44}} & +6.67 & \multicolumn{1}{c|}{+7.61}      & +3.18 & \multicolumn{1}{c|}{\bf{+3.93}} & +2.94 & \multicolumn{1}{c|}{+3.38}      & -1.13  & \multicolumn{1}{c|}{\bf{+4.00}} & +1.29  & \multicolumn{1}{c|}{\bf{+7.78}} & +1.55  & \multicolumn{1}{c|}{+1.09}      & +2.16 & +2.87      \\
                            & \multicolumn{1}{l|}{GaQR (S)}   & +3.77 & \multicolumn{1}{c|}{+3.80}      & +3.98 & \multicolumn{1}{c|}{+3.29}      & +1.63 & \multicolumn{1}{c|}{+1.66}      & +1.45 & \multicolumn{1}{c|}{\bf{+1.85}} & +0.01  & \multicolumn{1}{c|}{\bf{+1.79}} & +2.17  & \multicolumn{1}{c|}{\bf{+4.21}} & +2.91  & \multicolumn{1}{c|}{+1.47}      & +1.51 & +3.24      \\ \midrule
\multirow{4}{*}{Contriever} & \multicolumn{1}{l|}{No Exp} & \cellcolor[HTML]{C0C0C0}41.72 & \multicolumn{1}{c|}{\cellcolor[HTML]{C0C0C0}39.81}      & \cellcolor[HTML]{C0C0C0}47.75 & \multicolumn{1}{c|}{\cellcolor[HTML]{C0C0C0}42.31}      & \cellcolor[HTML]{C0C0C0}35.50 & \multicolumn{1}{c|}{\cellcolor[HTML]{C0C0C0}39.48}      & \cellcolor[HTML]{C0C0C0}30.89 & \multicolumn{1}{c|}{\cellcolor[HTML]{C0C0C0}32.33}      & \cellcolor[HTML]{C0C0C0}44.07  & \multicolumn{1}{c|}{\cellcolor[HTML]{C0C0C0}38.62}      & \cellcolor[HTML]{C0C0C0}87.88  & \multicolumn{1}{c|}{\cellcolor[HTML]{C0C0C0}89.83}      & \cellcolor[HTML]{C0C0C0}46.43  & \multicolumn{1}{c|}{\cellcolor[HTML]{C0C0C0}32.46}      & \cellcolor[HTML]{C0C0C0}69.57 & \cellcolor[HTML]{C0C0C0}56.14      \\
                            & \multicolumn{1}{l|}{Q2E}    & +1.47 & \multicolumn{1}{c|}{\bf{+4.31}} & +2.03 & \multicolumn{1}{c|}{\bf{+3.77}} & +1.11 & \multicolumn{1}{c|}{+0.90}      & +1.68 & \multicolumn{1}{c|}{+2.03}      & -8.63  & \multicolumn{1}{c|}{\bf{-4.38}} & -7.23  & \multicolumn{1}{c|}{\bf{-2.88}} & -7.48  & \multicolumn{1}{c|}{\bf{-0.65}} & -7.10 & \bf{+2.07} \\
                            & \multicolumn{1}{l|}{Q2D}    & -2.34 & \multicolumn{1}{c|}{\bf{+2.52}} & -0.24 & \multicolumn{1}{c|}{\bf{+3.36}} & -1.92 & \multicolumn{1}{c|}{\bf{+0.34}} & +1.03 & \multicolumn{1}{c|}{\bf{+2.60}} & -14.10 & \multicolumn{1}{c|}{\bf{-8.24}} & -13.44 & \multicolumn{1}{c|}{\bf{-6.89}} & -10.02 & \multicolumn{1}{c|}{\bf{-1.89}} & -9.48 & \bf{-0.10} \\
                            & \multicolumn{1}{l|}{GaQR (S)}   & +0.55 & \multicolumn{1}{c|}{\bf{+2.16}} & +0.11 & \multicolumn{1}{c|}{\bf{+1.18}} & +0.07 & \multicolumn{1}{c|}{\bf{+1.10}} & +0.14 & \multicolumn{1}{c|}{\bf{+1.14}} & -7.13  & \multicolumn{1}{c|}{\bf{-3.39}} & -6.55  & \multicolumn{1}{c|}{\bf{-1.95}} & -5.67  & \multicolumn{1}{c|}{\bf{-0.40}} & -7.43 & \bf{+0.10} \\ \bottomrule
\end{tabular}
  \label{tab:rq1_dataset}
  }
  \vspace{-1em}
\end{table*}

\subsection{Evaluation of LLM's Knowledge} 
\label{sec:llm_evaluation}
To evaluate the knowledge of LLMs regarding a given query,
we had an LLM produce a response and compared it with the query’s answer~\footnote{Prompts and settings are available at https://github.com/aken12/LLM-based-QE-fails}.
We took two different approaches to deciding the presence and absence of knowledge, depending on the length of answers in the datasets.
An exact-match approach was used for the datasets comprising short answers (i.e., NQ and TQ). 
An LLM is considered to have sufficient knowledge if its response to a given query exactly matches the answer to the query; otherwise, the LLM is considered to lack sufficient knowledge.
An LLM-based approach was used for the datasets containing longer answers (i.e., MS MARCO and BioASQ).
We had gpt-4-turbo-2024-04-09 evaluate the response from LLMs with prompts based on previous studies~\cite{wang2024evaluating,chiang2024chatbot}.
An LLM is considered to have sufficient knowledge if the evaluation result is either correct or partially correct.

\begin{table}[t]
  \caption{Improvements from different QE methods and LLMs used for QE (BM25 in MS MARCO).}
  \centering
  \scalebox{0.70}{%
    \begin{tabular}{llcccc}
\toprule
\multirow{2}{*}{LLM} & \multirow{2}{*}{Expansion}  & \multicolumn{2}{c}{NDCG@10}                                        & \multicolumn{2}{c}{Recall@100}                                     \\ \cmidrule{3-6} 
                     &                             & \raisebox{0.20ex}{\huge \emojiBox}                              & \raisebox{0.20ex}{\huge \emojiCheck}                                                       & \raisebox{0.20ex}{\huge \emojiBox}                              & \raisebox{0.20ex}{\huge \emojiCheck}                                  \\ [-1ex] \midrule \midrule
\multirow{4}{*}{GPT}                  & \multicolumn{1}{l|}{No Exp} & \cellcolor[HTML]{C0C0C0}24.76 & \multicolumn{1}{c|}{\cellcolor[HTML]{C0C0C0}22.01}      & \cellcolor[HTML]{C0C0C0}65.44 & \cellcolor[HTML]{C0C0C0}65.93      \\
                     & \multicolumn{1}{l|}{Q2E}    & -0.87 & \multicolumn{1}{c|}{\bf{+2.59}} & -1.05 & \bf{+4.84} \\
                     & \multicolumn{1}{l|}{Q2D}    & -0.55 & \multicolumn{1}{c|}{\bf{+5.20}} & +1.91 & \bf{+9.72} \\
                     & \multicolumn{1}{l|}{GaQR (T)}   & +0.07 & \multicolumn{1}{c|}{\bf{+1.81}} & +2.50 & \bf{+4.31} \\ \midrule
\multirow{4}{*}{Llama8b}              & \multicolumn{1}{l|}{No Exp} & \cellcolor[HTML]{C0C0C0}27.24 & \multicolumn{1}{c|}{\cellcolor[HTML]{C0C0C0}20.05}      & \cellcolor[HTML]{C0C0C0}68.01 & \cellcolor[HTML]{C0C0C0}64.38      \\
                     & \multicolumn{1}{l|}{Q2E}    & -1.60 & \multicolumn{1}{c|}{\bf{+1.59}} & -0.75 & \bf{+4.47} \\
                     & \multicolumn{1}{l|}{Q2D}    & -1.13 & \multicolumn{1}{c|}{\bf{+4.00}} & +1.29 & \bf{+7.78} \\
                     & \multicolumn{1}{l|}{GaQR (S)}   & +0.01 & \multicolumn{1}{c|}{\bf{+1.79}} & +2.17 & \bf{+4.21} \\ \bottomrule
\end{tabular}

  \label{tab:rq1_query_expansion_method}
  }
  \vspace{-1em}  
\end{table}

\section{Experimental Results} 

\label{sec:viewpoint_relation}

\subsection{Results for RQ1}
Table~\ref{tab:rq1_dataset} shows the improvements from QE at different knowledge levels.
``No Exp'' (gray) represents the results without QE, 
while the others indicate improvements over ``No Exp'' by each method.
Statistically significant differences among different knowledge levels (i.e., \raisebox{-0.05em}{\includegraphics[height=0.65em]{figures/white_large_square.pdf}} ~~ vs. \raisebox{-0.05em}{\includegraphics[height=0.65em]{figures/white_check_mark.pdf}}), identified by a t-test ($p\le0.05$), are highlighted in bold.
As a similar trend was observed for E5-base-v2, 
only Contriever's results were reported due to space limitations.
For the same reason, unless otherwise specified, we report the results for Llama8b.

The results show that when the LLM lacks knowledge about the query, the improvements from QE are significantly lower, especially in terms of NDCG@10, except for BM25 in BioASQ and a few other cases.
These results indicate that LLM-based QE loses effectiveness when the model is unfamiliar with the query, particularly in general domains rather than specialized ones like BioASQ.

Table~\ref{tab:rq1_query_expansion_method} presents the improvements from QE at different knowledge levels, focusing on differences in the QE methods and LLMs used for QE.
GPT and Llama8b show the same trend: when the LLM lacks knowledge about the query, 
LLM-based QE methods are significantly less effective.
It is worth mentioning that GaQR exhibits a distinctive trend compared to the other QE methods.
While Q2E and Q2D achieved larger improvements than GaQR when the LLM has sufficient query knowledge, GaQR is the most effective QE method when the LLM lacks sufficient knowledge.
This difference is likely due to GaQR's expansion strategy (see Section~\ref{sec:expansion_format}).
Q2E and Q2D enhance the retrieval performance by adding relevant keywords and answers to the original query, while GaQR paraphrases a given query for better retrieval effectiveness.
As paraphrasing may not require knowledge about the query when compared to generating keywords or answers, GaQR is considered less susceptible to the presence or absence of knowledge. 

Table~\ref{tab:rq1_retrieval_model} shows the improvements from QE at different knowledge levels, focusing on dense retrieval models.
Our results confirm an earlier finding~\cite{weller-etal-2024-generative} that high-performing retrievers tend to be adversely affected by QE: performance declines were observed in both retrieval models, except for Recall@100 for Q2E with Contriever.
However, we found that when the LLM has sufficient knowledge, the negative impact of QE is lower than when the LLM does not.

\begin{table}[t]
  \caption{
  Improvements from QE with different dense retrieval models (BioASQ).}
  \centering
  \scalebox{0.77}{%
    \begin{tabular}{llcccc}
\toprule
\multirow{2}{*}{Retrieval}  & \multicolumn{1}{c}{\multirow{2}{*}{Expansion}} & \multicolumn{2}{c}{NDCG@10}                                                              & \multicolumn{2}{c}{Recall@100}                                     \\ \cmidrule{3-6} 
                            & \multicolumn{1}{c}{}                           & \raisebox{0.20ex}{\huge \emojiBox}                              & \raisebox{0.20ex}{\huge \emojiCheck}                                                       & \raisebox{0.20ex}{\huge \emojiBox}                              & \raisebox{0.20ex}{\huge \emojiCheck}                                  \\[-1ex] \midrule \midrule
\multirow{3}{*}{Contriever} & \multicolumn{1}{l|}{No Exp}                    & \cellcolor[HTML]{C0C0C0}46.43  & \multicolumn{1}{c|}{\cellcolor[HTML]{C0C0C0}32.46}      & \cellcolor[HTML]{C0C0C0}69.57 & \cellcolor[HTML]{C0C0C0}56.14      \\
                            & \multicolumn{1}{l|}{Q2E}                       & -7.48  & \multicolumn{1}{c|}{\bf{-0.65}} & -7.10 & \bf{+2.07} \\
                            & \multicolumn{1}{l|}{Q2D}                       & -10.02 & \multicolumn{1}{c|}{\bf{-1.89}} & -9.48 & \bf{-0.10} \\ \midrule
\multirow{3}{*}{E5-base-v2} & \multicolumn{1}{l|}{No Exp}                    & \cellcolor[HTML]{C0C0C0}51.76  & \multicolumn{1}{c|}{\cellcolor[HTML]{C0C0C0}36.80}      & \cellcolor[HTML]{C0C0C0}72.44 & \cellcolor[HTML]{C0C0C0}57.70      \\
                            & \multicolumn{1}{l|}{Q2E}                       & -10.30 & \multicolumn{1}{c|}{\bf{-5.28}} & -8.19 & \bf{-0.06} \\
                            & \multicolumn{1}{l|}{Q2D}                       & -11.25 & \multicolumn{1}{c|}{\bf{-4.27}} & -9.93 & \bf{-0.71} \\ \bottomrule
\end{tabular}
  \label{tab:rq1_retrieval_model}
  }
  \vspace{-1em}  
\end{table}

\begin{figure}[t]
    \centering
    \scalebox{0.875}{%
        \includegraphics[width=0.5\textwidth]{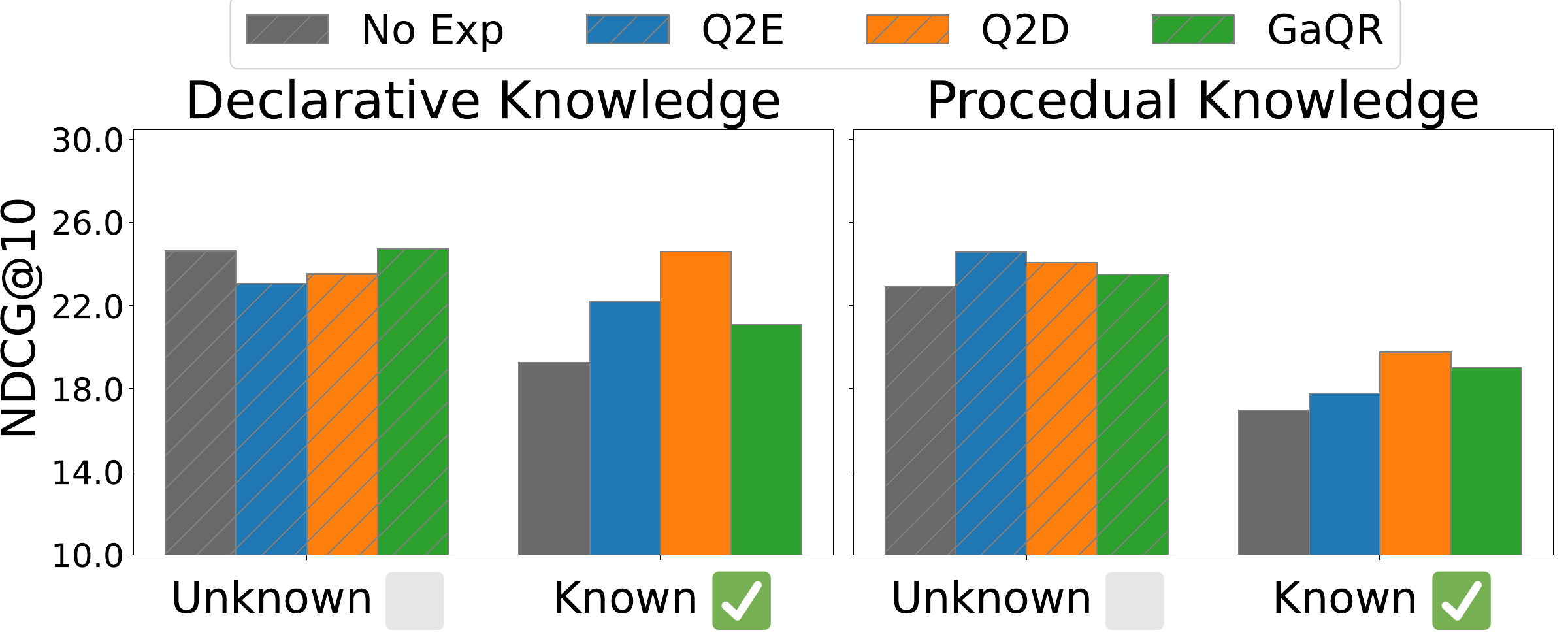}
    }
    \caption{QE performances for declarative and procedural knowledge queries.}
    \label{fig:RQ1_additional}
    \vspace{-1em}    
\end{figure}

\begin{table*}[t]
  \caption{
  Improvements from QE for queries with different ambiguity levels. Bold values denote statistically significant differences between different query ambiguity levels.
  }
  \centering
  \scalebox{0.7}{%
    \begin{tabular}{llcccccccccccc} 
\toprule
                            &                             & \multicolumn{4}{c}{AmbigDocs}                                                                                                                                                         & \multicolumn{4}{c}{TREC Web}                                                                                                                                                    & \multicolumn{4}{c}{AmbigQA}                                                                                                                                     \\ \cmidrule{3-14} 
\multirow{2}{*}{Retrieval}  & \multirow{2}{*}{Expansion}  & \multicolumn{2}{c}{NDCG@10}                                                               & \multicolumn{2}{c}{Recall@100}                                                            & \multicolumn{2}{c}{NDCG@10}                                                              & \multicolumn{2}{c}{Recall@100}                                                            & \multicolumn{2}{c}{NDCG@10}                                                              & \multicolumn{2}{c}{Recall@100}                                       \\ \cmidrule{3-14} 
                            &                             & \multicolumn{1}{c}{High}       & \multicolumn{1}{c}{Low}                                  & \multicolumn{1}{c}{High}       & \multicolumn{1}{c}{Low}                                  & \multicolumn{1}{c}{High}       & \multicolumn{1}{c}{Low}                                 & \multicolumn{1}{c}{High}       & \multicolumn{1}{c}{Low}                                  & \multicolumn{1}{c}{High}       & \multicolumn{1}{c}{Low}                                 & \multicolumn{1}{c}{High}       & \multicolumn{1}{c}{Low}             \\ \midrule \midrule
\multirow{4}{*}{BM25}       & \multicolumn{1}{l|}{No Exp} & \cellcolor[HTML]{C0C0C0}19.70  & \multicolumn{1}{l|}{\cellcolor[HTML]{C0C0C0}35.55}       & \cellcolor[HTML]{C0C0C0}60.01  & \multicolumn{1}{l|}{\cellcolor[HTML]{C0C0C0}76.94}       & \cellcolor[HTML]{C0C0C0}44.06  & \multicolumn{1}{l|}{\cellcolor[HTML]{C0C0C0}45.42}      & \cellcolor[HTML]{C0C0C0}58.78  & \multicolumn{1}{l|}{\cellcolor[HTML]{C0C0C0}64.70}       & \cellcolor[HTML]{C0C0C0}20.18  & \multicolumn{1}{l|}{\cellcolor[HTML]{C0C0C0}22.93}      & \cellcolor[HTML]{C0C0C0}41.64  & \cellcolor[HTML]{C0C0C0}49.90       \\
                            & \multicolumn{1}{l|}{Q2E}    & +1.09  & \multicolumn{1}{l|}{\bf{+2.44}}  & -0.49  & \multicolumn{1}{l|}{\bf{+0.86}}  & +11.31 & \multicolumn{1}{l|}{+5.08}      & -17.39 & \multicolumn{1}{l|}{\bf{-4.60}}  & +8.05  & \multicolumn{1}{l|}{+7.76}      & +9.86  & +11.78      \\
                            & \multicolumn{1}{l|}{Q2D}    & +2.61  & \multicolumn{1}{l|}{\bf{+4.19}}  & -0.02  & \multicolumn{1}{l|}{\bf{+1.78}}  & +13.58 & \multicolumn{1}{l|}{+13.54}     & -13.90 & \multicolumn{1}{l|}{\bf{-5.37}}  & +12.50 & \multicolumn{1}{l|}{+11.81}     & +13.14 & \bf{+16.64} \\
                            & \multicolumn{1}{l|}{GaQR (S)}   & +2.31  & \multicolumn{1}{l|}{\bf{+3.69}}  & +3.58  & \multicolumn{1}{l|}{+3.83}       & +12.45 & \multicolumn{1}{l|}{+10.97}     & -11.09 & \multicolumn{1}{l|}{-5.93}       & +4.82  & \multicolumn{1}{l|}{+3.98}      & +6.99  & +6.18       \\ \midrule
\multirow{4}{*}{Contriever} & \multicolumn{1}{l|}{No Exp} & \cellcolor[HTML]{C0C0C0}30.21  & \multicolumn{1}{l|}{\cellcolor[HTML]{C0C0C0}51.05}       & \cellcolor[HTML]{C0C0C0}72.84  & \multicolumn{1}{l|}{\cellcolor[HTML]{C0C0C0}88.71}       & \cellcolor[HTML]{C0C0C0}55.24  & \multicolumn{1}{l|}{\cellcolor[HTML]{C0C0C0}51.93}      & \cellcolor[HTML]{C0C0C0}55.69  & \multicolumn{1}{l|}{\cellcolor[HTML]{C0C0C0}61.07}       & \cellcolor[HTML]{C0C0C0}36.49  & \multicolumn{1}{l|}{\cellcolor[HTML]{C0C0C0}38.64}      & \cellcolor[HTML]{C0C0C0}63.68  & \cellcolor[HTML]{C0C0C0}70.95       \\
                            & \multicolumn{1}{l|}{Q2E}    & -8.07  & \multicolumn{1}{l|}{-9.14}       & -19.15 & \multicolumn{1}{l|}{\bf{-10.67}} & +0.18  & \multicolumn{1}{l|}{+2.78}      & -13.10 & \multicolumn{1}{l|}{\bf{-3.91}}  & +3.29  & \multicolumn{1}{l|}{+3.80}      & +3.92  & +5.71       \\
                            & \multicolumn{1}{l|}{Q2D}    & \bf{-10.61} & \multicolumn{1}{l|}{-13.93} & -25.52 & \multicolumn{1}{l|}{\bf{-17.00}} & +0.76  & \multicolumn{1}{l|}{+5.15}      & -8.03  & \multicolumn{1}{l|}{-4.03}       & +4.44  & \multicolumn{1}{l|}{+3.81}      & +5.39  & +6.34       \\
                            & \multicolumn{1}{l|}{GaQR (S)}   & \bf{-8.13}  & \multicolumn{1}{l|}{-9.37}  & -18.32 & \multicolumn{1}{l|}{\bf{-11.16}} & -3.47  & \multicolumn{1}{l|}{+0.34}      & -31.43 & \multicolumn{1}{l|}{\bf{-24.59}} & +2.66  & \multicolumn{1}{l|}{+1.86}      & +2.92  & +3.79       \\ \midrule
\multirow{4}{*}{E5-base-v2} & \multicolumn{1}{l|}{No Exp} & \cellcolor[HTML]{C0C0C0}31.17  & \multicolumn{1}{l|}{\cellcolor[HTML]{C0C0C0}52.98}       & \cellcolor[HTML]{C0C0C0}70.05  & \multicolumn{1}{l|}{\cellcolor[HTML]{C0C0C0}86.85}       & \cellcolor[HTML]{C0C0C0}57.31  & \multicolumn{1}{l|}{\cellcolor[HTML]{C0C0C0}55.02}      & \cellcolor[HTML]{C0C0C0}53.35  & \multicolumn{1}{l|}{\cellcolor[HTML]{C0C0C0}59.79}       & \cellcolor[HTML]{C0C0C0}47.61  & \multicolumn{1}{l|}{\cellcolor[HTML]{C0C0C0}49.43}      & \cellcolor[HTML]{C0C0C0}70.96  & \cellcolor[HTML]{C0C0C0}78.03       \\
                            & \multicolumn{1}{l|}{Q2E}    & \bf{-15.96} & \multicolumn{1}{l|}{-23.97} & -34.05 & \multicolumn{1}{l|}{\bf{-28.61}} & -7.95  & \multicolumn{1}{l|}{-3.35}      & -19.36 & \multicolumn{1}{l|}{\bf{-7.95}}  & -9.04  & \multicolumn{1}{l|}{\bf{-5.45}} & -8.11  & \bf{-4.29}  \\
                            & \multicolumn{1}{l|}{Q2D}    & \bf{-13.18} & \multicolumn{1}{l|}{-19.86} & -31.35 & \multicolumn{1}{l|}{\bf{-25.61}} & -2.43  & \multicolumn{1}{l|}{\bf{+3.31}} & -9.25  & \multicolumn{1}{l|}{\bf{-3.80}}  & -3.40  & \multicolumn{1}{l|}{-2.93}      & -1.96  & -0.01       \\ 
                               & \multicolumn{1}{l|}{GaQR (S)}   & \bf{-18.29} & \multicolumn{1}{l|}{-26.98} & -37.57 & \multicolumn{1}{l|}{\bf{-32.17}} & -6.59  & \multicolumn{1}{l|}{\bf{+2.13}} & -15.62 & \multicolumn{1}{l|}{\bf{-8.18}}  & -12.33 & \multicolumn{1}{l|}{-10.29}     & -13.38 & \bf{-9.69}  \\
                            \bottomrule
\end{tabular}
  }
  \label{tab:rq2_dataset}
  \vspace{-1em}
\end{table*}

\paragraph{Effect of query types.}
To clarify why QE effectiveness deteriorates when the LLM’s background knowledge is limited, we separately examined declarative and procedural queries, following the web-query taxonomy~\cite{eickhoff2014lessons}. Figure~\ref{fig:RQ1_additional} shows NDCG@10 of BM25 in MS MARCO, treating what/who queries as declarative and how queries as procedural~\cite{lupart2023ms}. In knowledge-limited situations, Q2E and Q2D failed to outperform ``No Exp'' on declarative queries, yet they still improved procedural queries. GaQR, by contrast, never fell below the baseline for either query type.

These patterns stem from differences in the underlying QE mechanisms. Q2D expands a query by generating a pseudo answer; for declarative queries, this answer often contains proper nouns, which retrieval models weight heavily. Incorrect proper nouns therefore degrade retrieval, making Q2D highly dependent on the LLM’s knowledge for declarative queries. Procedural queries seldom require proper nouns, so occasional inaccuracies in the expansion are less harmful. GaQR avoids answer generation altogether and is consequently less sensitive to the LLM’s knowledge availability.

\begin{figure}[t]
    \centering
    \scalebox{0.9}{%
        \includegraphics[width=0.5\textwidth]{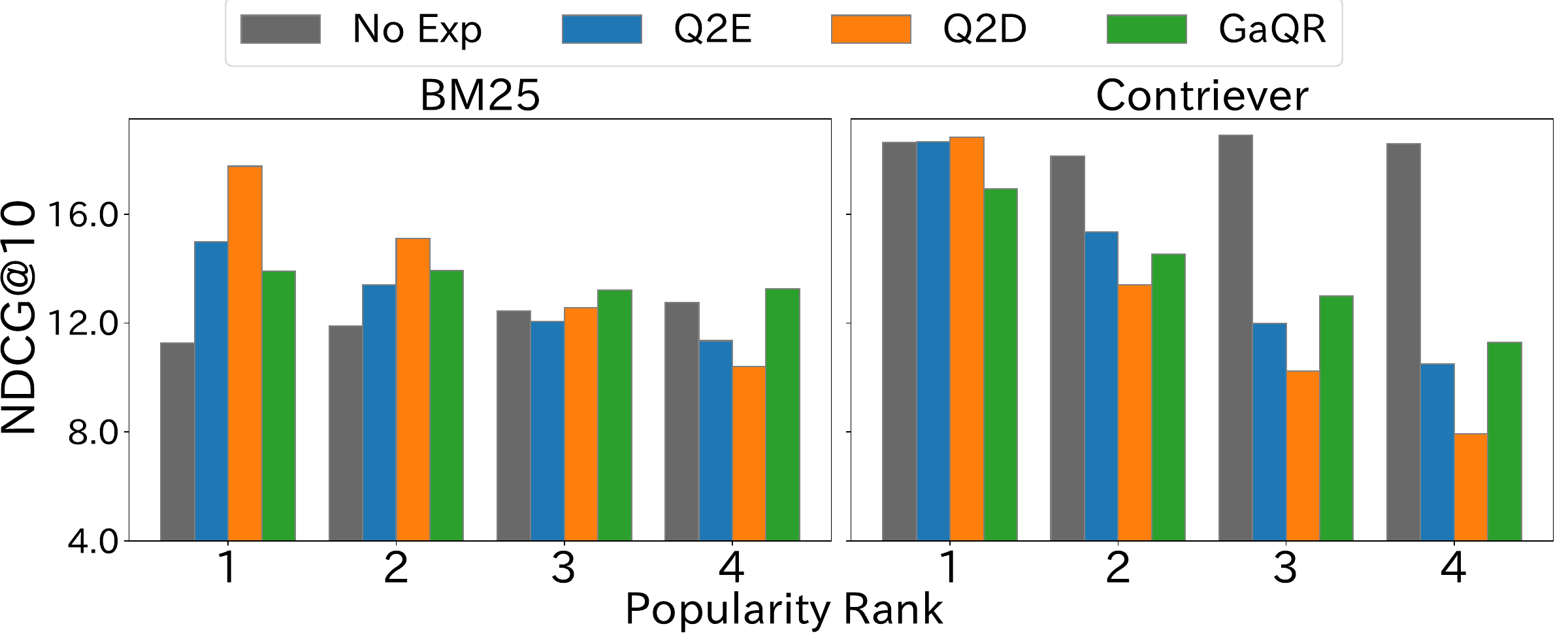}
    }
    \caption{Popularity bias analysis. ``1'' indicates the most popular query group, while ``4'' indicates the least popular query group.}
    \label{fig:RQ2_additional}
    \vspace{-1em}
\end{figure}

\subsection{Results for RQ2}
Table~\ref{tab:rq2_dataset} shows the improvements from QE for high- and low-ambiguity queries.
From Table~\ref{tab:rq2_dataset}, we can observe that the improvements in terms of Recall@100 are significantly smaller for high-ambiguity queries than low-ambiguity ones in most settings in AmbigDocs and TREC Web and in a few settings in AmbigQA.
The results are not conclusive in terms of NDCG@10: 
the improvements for high-ambiguity queries are smaller for BM25 in AmbigDocs and for E5-base-v2 in TREC Web,
while those for low-ambiguity queries are smaller for Contriever and E5-base-v2 in AmbigDocs.
A possible reason for an unclear trend in AmbigQA is the type of query ambiguity.
In AmbigQA~\cite{min2020ambigqa}, 
only 23\% of ambiguous queries are categorized into ``entity references'' (entities referred in a query are ambiguous).
Other queries are ambiguous because their answer types or time are ambiguous.
We hypothesize that relevant documents for different entity references can be highly diverse,
while those for different answer types or time are similar to each other. 
Thus AmbigQA likely offers less document diversity per query than datasets that focus on entity ambiguity, diminishing the gap between low- and high-ambiguity queries.

\paragraph{Analysis of popularity bias.}  \label{sec:pop_bias}
Since LLMs are known to privilege popular entities~\cite{he2023large}, we hypothesize that LLM-based QE is less effective for ambiguous queries because it primarily surfaces documents for only the most popular interpretations. To confirm this, we evaluated retrieval performance using only the documents relevant to the $x$-th most popular interpretation of each query, where popularity is measured by Wikipedia page views following PopQA~\cite{mallen2022not}. Figure~\ref{fig:RQ2_additional} shows results on AmbigDocs queries with at least four interpretations. QE generally improves, or at least does not harm, the ``No Exp'' baseline for highly popular interpretations; however, its effectiveness diminishes for rarer interpretations on both BM25 and Contriever. The sole exception is GaQR combined with BM25, which consistently outperforms the baseline regardless of popularity. We attribute GaQR’s robustness to its strategy of rewriting the query rather than generating answers or keyword lists, rendering it less sensitive to the popularity bias of the underlying LLM.

\section{Conclusion}
This paper systematically analyzed the failure patterns of LLM-based QE, focusing on two factors—LLM's knowledge of the query and query ambiguity.
Our findings reveal that when an LLM lacks knowledge about a query, LLM-based QE can significantly degrade retrieval performance, particularly in general domains. 
Moreover, we observed that highly ambiguous queries reduce recall, as our analysis suggests that LLM-based QE tends to favor popular interpretations of the query and overlook other interpretations.
Our future work is as follows: (i) test whether these issues persist with advanced pipelines (e.g., pseudo-relevance feedback, chain-of-thought prompting, multiple expansions) and (ii) develop a module that assesses LLM’s knowledge and query ambiguity for adaptive QE.

\begin{acks}
This work was supported by Japan Society for the Promotion of Science KAKENHI Grant Numbers JP23K28090 and JP23K25159. 
\end{acks}

\bibliographystyle{ACM-Reference-Format}
\balance
\bibliography{references}

\end{document}